\global\def\draftcontrol{0}

   \def\versionno{ universality2 }

\catcode`\@=11

\expandafter\ifx\csname draftcontrol\endcsname\relax\global\def\draftcontrol{0}
\fi

{\count255=\time\divide\count255 by 60
\xdef\hourmin{\number\count255}
\multiply\count255 by-60\advance\count255 by\time
\xdef\hourmin{\hourmin:\ifnum\count255<10 0\fi\the\count255}}
\def\draftdate{\number\month/\number\day/\number\year\ \ \ \hourmin }

\newcommand\makepapertitle{\par
  \begingroup
    \renewcommand\thefootnote{\@fnsymbol\c@footnote}%
    \def\@makefnmark{\rlap{\@textsuperscript{\normalfont\@thefnmark}}}%
    \long\def\@makefntext##1{\parindent 1em\noindent
            \hb@xt@1.8em{%
                \hss\@textsuperscript{\normalfont\@thefnmark}}##1}%
     \newpage
     \global\@topnum\z@   
     \@makepapertitle
     \thispagestyle{empty}\@thanks
  \endgroup
  \setcounter{footnote}{0}%
  \global\let\thanks\relax
  \global\let\makepapertitle\relax
  \global\let\@makepapertitle\relax
  \global\let\@thanks\@empty
  \global\let\@author\@empty
  \global\let\@date\@empty
  \global\let\@title\@empty
  \global\let\title\relax
  \global\let\author\relax
  \global\let\date\relax
  \global\let\and\relax
  \def\version{\let\version\@version\@gobble}
}
\def\@makepapertitle{%
  \newpage
   \ifnum\draftcontrol=1 {}
   \version\versionno
   \vskip 3em%
   \else
   \hfill\hbox to 3cm {\parbox{4cm}{\@pubnum}\hss}%
   \vskip 3em%
   \fi
   \begin{center}%
   \let \footnote \thanks
     {\LARGE {\@title}}%
     \vskip 1.5em%
     {\normalsize
       \lineskip .5em%
       \begin{tabular}[t]{c}%
         \@author
       \end{tabular}\par}%
     \vskip 1.5em%
     {\@bstract}%
     \end{center}%
     \vskip 1.5em
     \@date%
   \par
}

\gdef\@pubnum{}
\def\pubnum#1{%
  \gdef\@pubnum{#1}}

\gdef\@bstract{}
\def\Abstract#1{%
  \gdef\@bstract{%
   \parbox{\textwidth-0pc}{%
   \centerline{\bf Abstract}\penalty1000%
\kern.2cm%
\noindent
\renewcommand\baselinestretch{1.0}%
{#1}}}
}

\def\ps@paper{\let\@mkboth\@gobbletwo%
     \ifnum\draftcontrol=1
    \def\@oddfoot{\hbox to \textwidth{\tiny \versionno \hfil\tiny\draftdate}%
    \hskip -\textwidth \hbox to \textwidth{\hfil\rm\thepage\hfil}}%
     \else\def\@oddfoot{\hbox to \textwidth{\hfil\rm\thepage\hfil}}
     \fi
     \let\@evenfoot\@oddfoot
}

\def\body{\clearpage
          \pagestyle{paper}
    }

\def\@version#1{\ifnum\draftcontrol=1
\typeout{}\typeout{#1}\typeout{}
\vskip3mm\centerline{\hbox{\fbox{\normalsize{\tt DRAFT -- #1 -- }
                   {\draftdate}}}}\vskip3mm
\fi}
\let\version\@version
\long\def\eqlabel#1{\ifnum\draftcontrol=1
                    \tag@false  
                    \tag*{(\theequation) \hbox to -0.2cm{\hspace{0cm}\small{#1}\hss}}
                    \refstepcounter{equation}
                    \edef\@currentlabel{\theequation}
                    \ltx@label{#1}          
                    \else
                    \label{#1}
                    \fi
                    }
\let\st@bibitem\@bibitem
\let\st@lbibitem\@lbibitem
\ifnum\draftcontrol=1
  \def\@bibitem#1{%
    \st@bibitem{#1}\a@@label{#1}\ignorespaces}
  \def\@lbibitem[#1]#2{%
    \st@lbibitem[#1]{#2}\a@@label{#2}\ignorespaces}
  \def\a@@label#1{%
    \gdef\a@lab{\smash{\normalfont\small#1}}
    \ifvmode
      \if@inlabel
        \global\setbox\@labels\hbox{%
          \llap{\a@lab\let\a@lab\relax
                \kern\@totalleftmargin\kern\marginparsep}%
          \box\@labels}%
      \fi
    \fi}
\fi

\documentclass[12pt,letterpaper]{article}

\usepackage{amsmath,amssymb,array,calc,rotating,epsfig,psfrag}
\usepackage[nosort]{cite}

\ifnum\draftcontrol=1
\fi

\renewcommand\baselinestretch{1.25}
\renewcommand\section{\@startsection {section}{1}{\z@}%
                                   {-3.5ex \@plus -1ex \@minus -.2ex}%
                                   {2.3ex \@plus.2ex}%
                                   {\normalfont\large\bfseries}}
\renewcommand\subsection{\@startsection{subsection}{2}{\z@}%
                                   {-3.25ex\@plus -1ex \@minus -.2ex}%
                                   {1.5ex \@plus .2ex}%
                                   {\normalfont\normalsize\bfseries}}
\renewcommand\subsubsection{\@startsection{subsubsection}{3}{\z@}%
                                   {-3.25ex\@plus -1ex \@minus -.2ex}%
                                   {1.5ex \@plus .2ex}%
                                   {\normalfont\normalsize\it}}
\renewcommand\paragraph{\@startsection{paragraph}{4}{\z@}%
                                   {-3.25ex\@plus -1ex \@minus -.2ex}%
                                   {1.5ex \@plus .2ex}%
                                   {\normalfont\normalsize\bf}}

\numberwithin{equation}{section}

\def\ie{{\it i.e.}}

\def\revise#1       {\raisebox{-0em}{\rule{3pt}{1em}}%
                     \marginpar{\raisebox{.5em}{\vrule width3pt\
                     \vrule width0pt height 0pt depth0.5em
                     \hbox to 0cm{\hspace{0cm}{%
                     \parbox[t]{4em}{\raggedright\footnotesize{#1}}}\hss}}}}

\newcommand\nxt[1]  {\\\fnxt#1}

\def\cala         {{\cal A}}

\def\calf         {{\cal F}}

\def\calk         {{\cal K}}
\def\call         {{\cal L}}
\def\calm         {{\cal M}}
\def\caln         {{\cal N}}
\def\calo         {{\cal O}}

\def\calv         {{\cal V}}

\def\reals        {{\mathbb R}}

\def\del          {\partial}

\def\sqr#1#2{{\vcenter{\vbox{\hrule height.#2pt
 \hbox{\vrule width.#2pt height#1pt \kern#1pt
 \vrule width.#2pt}\hrule height.#2pt}}}}

\newcommand{\fft}[2]{{\frac{#1}{#2}}}

\def\w{\omega}
\def\a{\alpha}
\def\p{\psi}
\def\b{\beta}
\def\rh{r_{horizon}}
\def\r{\rho}
\def\l{\lambda}

\catcode`\@=12

\begin{document}


\title{The shear viscosity of gauge theory plasma with chemical potentials}

\pubnum{%
UWO-TH-06/19}
\date{October 2006}

\author{
Paolo Benincasa$^{1}$, Alex Buchel$^{1,2}$ and Roman Naryshkin$^{1,3}$\\[0.4cm]
\it $^1$Department of Applied Mathematics\\
\it University of Western Ontario\\
\it London, Ontario N6A 5B7, Canada\\[0.1cm]
\it $^2$Perimeter Institute for Theoretical Physics\\
\it Waterloo, Ontario N2J 2W9, Canada\\
\it $^3$Physics Department\\
\it Taras Shevchenko Kiev National University\\
\it Prosp.Glushkova 6, Kiev 03022, Ukraine\\
}

\Abstract{
We consider strongly coupled gauge theory plasma with conserved global
charges that allow for a dual gravitational description. We study the
shear viscosity of the gauge theory plasma in the presence of chemical
potentials for these charges.  Using gauge theory/string theory
correspondence we prove that at large 't Hooft coupling the ratio of
the shear viscosity to the entropy density is universal.
}


\makepapertitle

\body

\version\versionno

\section{Introduction}
The gauge theory/string theory correspondence of Maldacena
\cite{m9711,m2} provides a valuable insight into the nonperturbative
dynamics of strongly coupled gauge theory plasma.  Originally, the 
string theory correspondence was formulated for static 
properties of strongly coupled gauge theories. It was pointed out  
in \cite{s1} that computing equilibrium two-point correlation
functions of  stress-energy tensor 
  one can extract the physics of the near-equilibrium (hydrodynamic)
description of strongly coupled hot gauge theory plasma. The computations of the 
finite temperature transport properties of the $\caln=4$ $SU(N)$ 
supersymmetric Yang-Mills (SYM) 
theory plasma  at large t' Hooft coupling\footnote{Finite 't Hooft coupling corrections to 
$\caln=4$ hydrodynamics were discussed in \cite{b1,b2}.} \cite{s1,s2}
were extended to various non-conformal  gauge theory plasmas in \cite{h1,h2,b3,aby,s3,b4,b5,b6,b7}.
One of the most surprising results of these computations was the discovery of the 
universality of the 
shear viscosity of strongly coupled gauge theory plasma. Specifically, for all gauge theories 
which allow for a dual supergravity description, and without chemical potentials for conserved 
global charges (if such charges are present),  it was shown that  the ratio of the shear viscosity 
$\eta$ to the entropy density $s$ is a universal constant \cite{u1,u2,u3}
\begin{equation}
\frac{\eta}{s}=\frac{1}{4\pi}\,,
\eqlabel{u}
\end{equation}
in units $\hbar=k_{B}=1$.

$\caln=4$ SYM  has a maximum $U(1)^3\subset SO(6)_R$ abelian subgroup of the R-symmetry for 
which one can introduce (at most three) different chemical potentials. At strong coupling and finite temperature, 
the dual supergravity description of this gauge theory plasma is given by a system of near extremal 
D3 branes with generically three different angular momenta along the five-sphere $S^5$ 
\cite{c1}\footnote{The misprints in  the five-form expression for this supergravity solution were 
fixed in \cite{c2}.}.
This supergravity solution allows for a consistent  Kaluza-Klein reduction to 
five dimensions, where it is described within $\caln=2$ supergravity  coupled to  
two abelian vectors commonly referred to as the STU model \cite{stu}. Within this 
effective five dimensional description, finite temperature $\caln=4$ SYM plasma is dual to 
non-extremal  black holes carrying generically three different $U(1)$ charges 
corresponding to three different $S^5$ angular momenta of the near-extremal D3 
branes in the type IIB supergravity description. Even though $\caln=4$ SYM plasma 
with chemical potentials violates the assumptions of the universality theorem \cite{u1,u2,u3}, 
explicit computation of  shear viscosity leads to \eqref{u} 
\cite{a1,a2,a3}. 

Expectation that \eqref{u} might in fact be more universal than anticipated  in \cite{u1,u2,u3} was 
strengthened\footnote{Shear viscosity of the M2-brane plasma also appears to satisfy the 
universal relation \eqref{u} \cite{a4}.}
by explicit construction of new models of  finite temperature 
gauge/string theory correspondence with an $R$-charge chemical potential 
\cite{bl}. Specifically, it was found  that shear viscosity of all  strongly coupled 
$Y^{p,q}$ quiver gauge theory \cite{y1,y2,y3,y4} plasmas with a $U(1)$ R-charge chemical 
potential satisfies the universal relation \eqref{u}. 
It was further conjectured in \cite{bl} that \eqref{u} holds true even in the presence of 
chemical potentials. 

In this paper we prove that \eqref{u} is indeed satisfied for generic strongly coupled gauge theory 
plasma with global charge chemical potentials. We begin with spelling out explicit 
assumptions (and their implications)  under which we obtain \eqref{u}.  
\nxt First, we consider strongly coupled gauge theory plasma that allow for a dual string theory 
description.  This allows us to use the gauge theory/string theory Minkowski-space 
prescription  \cite{ss} for computing two-point correlators of the stress-energy tensor.
We will work in the regime of large 't Hooft coupling, where 
the supergravity approximation to string theory is valid.  
\nxt The gauge theory/string theory correspondence relates a 
$(D+1)$-dimensional 
strongly coupled gauge theory on $\reals^{D,1}$ space-time 
to a particular background of ten-dimensional type IIB supergravity. 
In all known examples of gauge theory/string theory correspondence it is possible to do a Kaluza-Klein 
reduction along a compact $(8-D)$-dimensional manifold and relate $(D+1)$-dimensional 
strongly coupled gauge theory to an effective $(D+2)$-dimensional gauged supergravity, obtained from a
consistent truncation of the full ten-dimensional supergravity. However,
we do not believe that this is 
always possible to do. In fact, the strongest form of the universality \eqref{u} proven in 
\cite{u3} does not rely on this assumption\footnote{Both proofs \cite{u1,u2} implicitly assume 
the existence of a consistent truncation. The proof \cite{u2} further implicitly 
assumes that the background is asymptotically flat, as it relies on the universality 
of the cross section for the graviton scattering from the black hole horizon \cite{sc1,sc2}. 
The latter universality was  derived only for asymptotically flat space-times, and 
in general it is not known how to ``re-attach'' the asymptotically flat region in gravitational 
dual to non-conformal gauge theories (other than those described by flat Dp branes).}.      
In this paper we assume that our gauge theory plasma is described by a type IIB gravitational 
background that allow for a consistent truncation to a $(D+2)$-dimensional gauged supergravity.
However, we will not assume that the geometry is  asymptotically flat, or that the 
asymptotically flat region can be ``re-attached''. Thus, our universality class of \eqref{u}
is necessarily weaker than that of the zero chemical potential case considered in \cite{u3}. 
Since we would like to study gauge theory plasma with finite chemical potentials, these gauge theories 
must have a set of abelian conserved charges --- otherwise we simply would not be able to 
introduce a chemical potential. In the dual supergravity picture, for each conserved $U(1)$ charge,
there must be a corresponding $U(1)$ isometry. Introducing a chemical potential for a particular 
$U(1)$ charge on a gauge theory side results in gauging corresponding $U(1)$ isometry on the 
supergravity side. In the effective $(D+2)$-dimensional gravitational description, besides  
Einstein-Hilbert term and  scalar fields,  one gets a set of  Maxwell fields, one for each gauged isometry 
(or a chemical potential from the gauge theory perspective). To summarize, we consider strongly coupled 
$(D+1)$-dimensional gauge theory plasmas which have a following dual effective $(D+2)$-dimensional 
gravitational description   
\begin{equation}
\begin{split}
S_{D+2}=\frac{1}{16\pi G_{D+2}}\int_{\calm_{D+2}} d\xi^{D+2}\sqrt{-g}\biggl\{R-\calk_{\a\b}(\phi)
\del_\mu\phi^\a\del^\mu\phi^\b-\tau_{ab}(\phi)F_{\mu\nu}^{(a)}F^{\mu\nu(b)}-\calv(\phi)\biggr\}\,,
\end{split}
\eqlabel{action}
\end{equation}
where $\calv$, $\calk_{\a\b}$ and $\tau_{ab}$ are arbitrary functions 
of scalar fields $\phi^{\a}$, and the index $(a)$ in 
Maxwell fields $F_{\mu\nu}^{(a)}=\del_\mu A_\nu^{(a)}-\del_{\nu} A_{\mu}^{(a)}$ 
runs over the set of nonzero chemical potentials.
Corresponding to finite temperature $(D+1)$-dimensional gauge theory with chemical potentials,
the effective action \eqref{action} must admit a  black $D$-dimensional brane solution, electrically 
charged under vector fields $A_{\mu}^{a}$. The $SO(D)$ invariance of the solution implies that 
it is of the form
\begin{equation}
\begin{split}
ds_{D+2}^2=&-c_1^{2}(r)\ (dt)^2+ c_2^2(r)\ \sum_{i=1}^D (dx^i)^2+c_3^2(r)\ (dr)^2\,,\\
A_{\mu}^{(a)}=&\delta_{\mu}^t\ \Phi^{(a)}(r)\,,\qquad \phi^{\a}=\phi^{\a}(r)\,,
\end{split}
\eqlabel{back}
\end{equation}
for some scalar potentials $\Phi^{(a)}$. 
\nxt Finally, we assume that the black brane horizon of \eqref{back} is nonsingular. This implies 
that there is a choice of the radial coordinate such that as $r\to r_{horizon}$
\begin{equation}
c_1\to \a_1 (r-\rh)^{1/2}\,,\qquad  c_2\to \a_2\,,\qquad  c_3\to \a_3 (r-\rh)^{-1/2}\,,
\eqlabel{hass}
\end{equation} 
where constants $\a_i$ satisfy
\begin{equation}
\a_1\ \a_2\ \a_3\ \ne 0\,.
\eqlabel{const}
\end{equation}
Notice that given \eqref{hass} the black brane temperature $T$ and the entropy density $s$ are correspondingly
\begin{equation}
T=\frac{\a_1}{4\pi \a_3}\,,\qquad s=\frac{\a_2^D}{4 G_{D+2}}\,.
\eqlabel{ts}
\end{equation}

\section{The proof}
In this section, we examine the hydrodynamics of the gauge theory 
plasma at finite chemical potential dual to the generic  black hole solution 
\eqref{back}.  In particular, using
prescription \cite{ss}, we compute the retarded Green's function of the
boundary stress-energy tensor $T_{\mu\nu}(t,x^i)$ ($\mu=\{t,x^i\}$) at
zero spatial momentum, and in the low-energy limit $\w\to 0$:
\begin{equation}
G^{R}_{12,12}(\w,0)=-i\int dt\,d^Dx\ e^{i\w t}\theta(t)
\langle[T_{12}(t,x^i),T_{12}(0,0)]\rangle\,.
\eqlabel{defgr}
\end{equation}
Computation of this Green's function allows for a determination of the
shear viscosity $\eta$ through the Kubo relation
\begin{equation}
\eta=\lim_{\w\to 0}\frac{1}{2\w i}\left[G^A_{12,12}(\w,0)-G^R_{12,12}(\w,0)
\right]\,,
\eqlabel{kubo}
\end{equation}
where the advanced Green's function is given by
$G^A(\w,0)=\left(G^R(\w,0)\right)^\ast$.
We find
\begin{equation}
G^{R}_{12,12}(\w,0)=-\frac{i\w s}{4\pi}\left(1+\calo\left(\frac{\w}{T}\right)
\right)\,,
\eqlabel{result}
\end{equation}
where $s$ is the entropy density.
Inserting this expression into (\ref{kubo}) yields the universal ratio \eqref{u}.

We begin the computation of \eqref{defgr} by recalling that the coupling
between the boundary value of the graviton and the stress-energy tensor of
a gauge theory is given by $\delta g_2^1T_1^2/2$. According to the
gauge/gravity prescription, in order to compute the retarded thermal
two-point function \eqref{defgr}, we should add a small bulk perturbation
$\delta g_{12}(t,r)$ to the metric \eqref{back}
\begin{equation}
ds_{D+2}^2\to ds_{D+2}^2+\delta g_{12}(t,r)\ dx^1dx^2\,, 
\eqlabel{metper}
\end{equation}
 and compute the on-shell
action as a functional of its boundary value $\delta g_{12}^b(t)$. Symmetry
arguments \cite{ne2} guarantee that for a perturbation of this type in the 
background \eqref{back} all the
other components of a generic perturbation $\delta g_{\mu\nu}$, along with the
gauge potentials perturbations $\delta A_\mu^{(a)}$  and scalar perturbations 
$\delta \phi^{\a}$, can be consistently set to zero.

Instead of working directly with $\delta g_{12}$, we find it convenient to
introduce the field $\p=\p(t,r)$ according to
\begin{equation}
\p=\frac 12 g^{11}\,\delta g_{12}=\frac 12 c_2^{-2}\ \delta g_{12}\,.
\eqlabel{defp}
\end{equation}
The retarded correlation function $G^R_{12,12}(\w,0)$ can be extracted
from the (quadratic) boundary effective action $S_{boundary}$ for the metric
fluctuations $\p^b$ given by
\begin{equation}
S_{boundary}[\p^b]=\int \frac{d^{D+1}k}{(2\pi)^{D+1}}\,\p^{b}(-\w)\,\calf(\w,r)\,
\p^{b}(\w)\bigg|^{\del\calm_{D+2}}_{horizon}\,,
\eqlabel{sssb}
\end{equation}
where
\begin{equation}
\p^b(\w)=\int \frac{d^{D+1}k}{(2\pi)^{D+1}} e^{-i\w t}\
\p(t,r)\bigg|_{\del\calm_{D+2}}\,.
\eqlabel{pb}
\end{equation}
In particular, the Green's function is given simply by
\begin{equation}
G^R_{12,12}(\w,0)=\lim_{\del\calm_{D+2}^r \to\del\calm_{D+2}}\ 2\ \calf^r(\w,r)\,,
\eqlabel{Gr}
\end{equation}
where $\mathcal F$ is the kernel in (\ref{sssb}).
The boundary metric functional is defined as
\begin{equation}
S_{boundary}[\p^b]=\lim_{\del\calm_{D+2}^r \to\del\calm_{D+2}}\biggl(
S^r_{bulk}[\p]+S_{GH}[\p]+S^{counter}[\p]\biggr)\,,
\eqlabel{bounfu}
\end{equation}
where $S^r_{bulk}$ is the bulk Minkowski-space effective supergravity action
\eqref{action} on a cut-off space $\calm_{D+2}^r$ (where $\calm_{D+2}$ in
\eqref{back} is regularized by the compact manifold $\calm_{D+2}^r$ with a
boundary $\del\calm_{D+2}^r$).  Also, $S_{GH}$ is the standard Gibbons-Hawking
term over the regularized boundary $\del\calm_{D+2}^r$.  The regularized bulk
action $S^r_{bulk}$ is evaluated on-shell for the bulk metric fluctuations
$\p(t,r)$ subject to the following boundary conditions:
\begin{equation}
\begin{split}
&(a):\quad\lim_{\del\calm_{D+2}^r \to\del\calm_{D+2}} \p(t,r)=\p^b(t)\,,\\
&(b):\quad\p(t,r)\ \hbox{is an incoming wave at the horizon}\,.
\end{split}
\eqlabel{bc}
\end{equation}
The purpose of the boundary counterterm $S^{counter}$ is to remove
divergent (as $\del\calm_{D+2}^r\to\del\calm_{D+2}$) and $\w$-independent
contributions from the kernel $\calf$ of \eqref{sssb}.

The effective bulk action for $\p(t,r)$ is derived in Appendix \ref{eom}.
It takes the form
\begin{equation}
\begin{split}
S_{bulk}[\p]\equiv&\frac{1}{16\pi G_{D+2}}\int_{\calm_{D+2}} d^{D+2}\xi\ \call_{D+2}=
\frac{1}{16\pi G_{D+2}}\int_{\calm_{D+2}} d^{D+2}\xi\ \biggl[\\
&c_1c_2^Dc_3
\biggl\{\frac{1}{2c_1^2} \left(\del_t \p\right)^2-\frac{1}{2c_3^2}
\left(\del_r\p\right)^2\biggr\}\\
&+\biggl\{-\del_t\left(\frac{2c_2^Dc_3}{c_1}\ \p\del_t\p\right)
+\del_r\left(
\frac{2c_2^Dc_1}{c_3}\ \p\del_r\p+\frac{c_1c_2^{D-1}c_2'}{c_3}\,\p^2
\right)
\biggr\}
\biggr]\,.
\end{split}
\eqlabel{ac2}
\end{equation}
The second line in \eqref{ac2} is the effective action for a minimally coupled
scalar in the geometry \eqref{back}, while the third line is a total
derivative. Thus the bulk equation of motion for $\p$ is that of a minimally
coupled scalar in \eqref{back}. Decomposing $\p$ as
\begin{equation}
\p(t,r)=e^{-i\w t}\p_\w(r)\,,
\eqlabel{decom}
\end{equation}
we find that the equation of motion reduces to
\begin{equation}\eqlabel{eomp}
\frac{\w^2\psi_\w}{c_{1}^{2}} + 
\frac{\psi_\w''}{c_{3}^{2}} +
\left[\ln{\frac{c_{1}c_{2}^{D}}{c_{3}}}\right]'\frac{\psi_\w'}{c_{3}^{2}}=0\,,
\end{equation}
where primes denote derivatives with respect to $r$.

Let's understand the characteristic indices of $\p_\w$ as $r\to \rh$. 
Consider the following ansatz  
\begin{equation}
\p_\w(r)\sim (\Lambda(r-\rh))^\l\,,
\eqlabel{ci}
\end{equation}
where $\Lambda$ is a characteristic, finite, $\w$-independent energy scale associated with
the background geometry \eqref{back}. For background geometries dual to conformal gauge theory
plasmas the role of $\Lambda$ is being played by the temperature. More generally, 
$\Lambda$ could be a scale at which conformal invariance is broken (either explicitly or 
spontaneously). It is crucial for $\Lambda$ to be finite  --- otherwise 
the hydrodynamic approximation to a strongly coupled gauge theory plasma would not 
applicable\footnote{This applicability of the hydrodynamic regime is an implicit 
assumption of all the studies of strongly coupled gauge theory plasma within gauge theory/string theory 
correspondence.}.  
 Substituting \eqref{ci} into \eqref{eomp}
and using the near horizon asymptotics \eqref{hass}, we find to leading order in $(\Lambda(r-\rh))\ll 1$,
\begin{equation}
(r-\rh)^{\l-1}\ \bigg\{\frac{\w^2}{\a_1^2}+\frac{\l(\l-1)}{\a_3^2}+\frac{\l}{\a_3^2}\biggr\}=0\,,
\end{equation} 
or 
\begin{equation}
\l=\pm i\ \frac{\a_3\w}{\a_1}=\pm i\ \frac{\w}{4\pi T}\,,
\end{equation}
where we used \eqref{ts}. Given \eqref{decom}, for the incoming wave at the horizon we 
must have 
\begin{equation}
\p_\w(r)^{incoming}\sim (\Lambda(r-\rh))^{-i\w/(4\pi T)}\,,\qquad  (\Lambda(r-\rh))\ll 1\,,
\eqlabel{pinc}
\end{equation} 
which to leading order in $\frac{\w}{T}$ takes the form
\begin{equation}
\p_\w(r)^{incoming}\sim 1-i \frac{\w}{4\pi T}\ln((\Lambda(r-\rh))\,,
\eqlabel{pinc1}
\end{equation} 
valid when both inequalities are satisfied
\begin{equation}
\bigg|\ln((\Lambda(r-\rh))\bigg|\gg 1\qquad \&\qquad \frac{\w}{T}\ \bigg|\ln((\Lambda(r-\rh))\bigg|\ll 1\,.
\eqlabel{overlap}
\end{equation}
Clearly, for finite $\Lambda$ and sufficiently small $\w$ there is an overlap region 
in \eqref{overlap}.

In what follows we will need a solution to \eqref{eomp}, subject to \eqref{bc}, 
to linear order in $\frac{\w}{T}$, \ie, in the low-frequency approximation. In this approximation 
one can neglect the first term in \eqref{eomp} and write down the most general solution 
as 
\begin{equation}
\p_\w(r)=\cala_1(\w)+\cala_2(\w)\ \int_r^{\infty} \frac{c_3(\r)d\r}{c_1(\r)c_2(\r)^D}\,,
\eqlabel{gensol}
\end{equation}  
where $\cala_i(\w)$ are constants, depending at most linearly on $\frac{\w}{T}$.
The first boundary condition  in \eqref{bc} implies that 
\begin{equation}
\cala_1(\w)=1\,.
\eqlabel{a1}
\end{equation}
Substituting  \eqref{hass} into \eqref{gensol} we obtain  the leading  near horizon 
behavior of  $\p_\w$
\begin{equation}
\p_\w(r)\approx 1- \frac{\a_3\cala_2(\w)}{\a_1\a_2^D}\ \ln(\Lambda(r-\rh))\,, 
\qquad \bigg|\ln(\Lambda(r-\rh))\bigg|\gg 1\,.
\eqlabel{nhb}
\end{equation}
Comparing \eqref{nhb} with  \eqref{pinc1} in the hydrodynamic approximation, \ie, for sufficiently 
small $\frac{\w}{T}$, we conclude that 
\begin{equation}
\cala_2(\w)=i\ \a_2^D\ \w\,.
\eqlabel{a2det}
\end{equation}
To summarize, we have to leading order in $\frac{\w}{T}$
\begin{equation}
\p_\w(r)=1+i\ \a_2^D\w\ \ \int_r^{\infty} \frac{c_3(\r)d\r}{c_1(\r)c_2(\r)^D}\,.
\eqlabel{gensol1}
\end{equation}  
Notice that as $r\to \infty$
\begin{equation}
\psi_\w(r)\to 1\,,\qquad \del_r\psi_\w\to -i\ \a_2^D\w\ \frac{c_3(r)}{c_1(r)c_2(r)^D}\,.
\eqlabel{lim}
\end{equation}

Once the bulk fluctuations are on-shell (\ie,\ satisfy equations of motion)
the bulk gravitational Lagrangian becomes a total derivative. From \eqref{ac2}
we find (without dropping any terms)
\begin{equation}
\call_{D+2}=\del_t J^t+\del_r J^r\,,
\eqlabel{tder}
\end{equation}
where
\begin{equation}
\begin{split}
J^t=&-\frac{3c_2^Dc_3}{2c_1}\,\p\del_t\p\,,\\
J^r=&\frac{3c_2^Dc_1}{2c_3}\,\p\del_r\p+\frac{c_2^{D-1}c_1c_2'}{c_3}\,\p^2\,.
\end{split}
\eqlabel{jr}
\end{equation}
Additionally, the Gibbons-Hawking term provides an extra contribution so that
\begin{equation}
J^r\to J^r-\frac{2c_2^Dc_1}{c_3}\,\p\del_r\p-\frac{(c_1c_2^D)'}{c_3}\,\p^2.
\eqlabel{ghshift}
\end{equation}

We are now ready to extract the kernel $\calf$ of \eqref{sssb}.
The regularized boundary effective action for $\p$ is
\begin{equation}
\begin{split}
S_{boundary}[\p]^r
=&\frac{1}{16\pi G_{D+2}}\int_{\del\calm^r_{D+2}}dt\,d^Dx\,
\biggl(-\frac{c_2^Dc_1}{2c_3}\,\p\del_r\p\biggr)+c.t.\,,
\end{split}
\eqlabel{breg}
\end{equation}
where, as prescribed in \cite{ss}, we need only to keep the boundary contribution.
In \eqref{breg} $c.t.$ stands for (finite) contact terms that will not be
important for computations.   
Substituting \eqref{decom}, \eqref{gensol1} into \eqref{breg}, and using \eqref{lim}, we
obtain $\calf(\w,r)$ in the limit $r\to \infty$
\begin{equation}
\begin{split}
\lim_{r\to \infty}
\calf^r(\w,r)=&-\frac{i\a_2^D\w}{32\pi G_{D+2}}\left(1+\calo\left(\frac{\w}{T}\right)\right)
\\
\approx& -\frac{i\w s}{8\pi}\,,
\end{split}
\eqlabel{regk}
\end{equation}
where we have recalled the definition of $s$ in \eqref{ts}.
Using \eqref{Gr} to extract the
Green's function from $\calf^r$ we find
\begin{equation}
G_{12,12}^R(\omega,0)\approx-\fft{i\w s}{4\pi}\,,
\end{equation}
at least in the low frequency limit $\omega\to0$.  This is the result
claimed in \eqref{result}, giving rise to the universal ratio of
shear viscosity to entropy density \eqref{u}.

\section*{Acknowledgments}
PB would like to thank the organizers of the ESF Cargese Summer
School 2006 {\it ``Strings and Branes: the Present Paradigm for Gauge
Interactions and Cosmology''} and the organizers of the SNFT 2006
Parma School of Theoretical Physics for the stimulating environment
created, as well as to all the participants at these schools without
whom they would have never been the same.  AB's research at Perimeter
Institute is supported in part by the Government of Canada through
NSERC and by the Province of Ontario through MEDT.  AB gratefully
acknowledges further support by an NSERC Discovery grant.

\appendix
\section{Effective bulk action for $\p$}\label{eom}

We begin with collecting some useful expressions. 
The trace of the Einstein equations derived from \eqref{action} takes the form
\begin{equation}
R=\calk_{\a\b}\del\phi^\a\del\phi^b+\frac{D-2}{D}\ \tau_{ab}F^{(a)}_{\mu\nu}F^{\mu\nu(b)}+\frac{D+2}{D}\ \calv
\,.
\eqlabel{trace}
\end{equation}
Additionally  we have 
\begin{equation}
\begin{split}
\sqrt{-g^{(0)}}\ R_{\ x_1}^{x_1(0)}=&-\left[\frac{c_1c_2^{D-1}c_2'}{c_3}\right]'
=\frac 1D\ \sqrt{-g^{(0)}} \biggl(\calv-\tau_{ab}F^{(a)}_{\mu\nu}F^{\mu\nu(b)}\biggr)^{(0)}\,,
\end{split}
\eqlabel{r11}
\end{equation}
where we used superscript $(0)$ to emphasize that the Ricci tensor component and the metric determinant  
are evaluated on the background \eqref{back}. 
The second equality in \eqref{r11} makes use of the Einstein equation for $ R_{\ x_1 x_1}^{(0)}$.

Expanding the effective action 
\eqref{action} to quadratic order in perturbation \eqref{metper}  and using \eqref{trace}, we find 
\begin{equation}
\begin{split}
S_{bulk}[\p]=&\frac{1}{16\pi G_{D+2}}\int_{\calm_{D+2}} d^{D+2}\xi \sqrt{-g^{(0)}}\left\{R^{(2)}-\frac 12\p^2\ \frac 2D 
\left(\calv-\tau_{ab}F^{(a)}_{\mu\nu}F^{\mu\nu(b)}\right)^{(0)}\right\}\\
=&\frac{1}{16\pi G_{D+2}}\int_{\calm_{D+2}} d^{D+2}\xi \sqrt{-g^{(0)}}\left\{R^{(2)}-\p^2\ R_{\ x_1}^{x_1(0)}\right\}\,,
\end{split}
\eqlabel{sb1}
\end{equation}
where in the second line we used identity \eqref{r11}.
In \eqref{sb1}, $R^{(2)}$ is a perturbation of the Ricci scalar, quadratic in $\p$ and its derivatives
\begin{equation}
\begin{split}
\sqrt{-g^{(0)}}R^{(2)}=&\sqrt{-g^{(0)}}\biggl\{\frac{1}{2c_1^2} \left(\del_t \p\right)^2-\frac{1}{2c_3^2}
\left(\del_r\p\right)^2\biggr\}\\
&+\biggl\{-\del_t\left(\frac{2c_2^Dc_3}{c_1}\ \p\del_t\p\right)
+\del_r\left(
\frac{2c_2^Dc_1}{c_3}\ \p\del_r\p+\frac{c_1c_2^{D-1}c_2'}{c_3}\,\p^2
\right)
\biggr\}\\
&+\sqrt{-g^{(0)}}\ \p^2\ R_{\ x_1}^{x_1(0)}\,,
\end{split}
\eqlabel{r2pert}
\end{equation}
where we again used \eqref{r11}. From \eqref{sb1} and \eqref{r2pert} we obtain \eqref{ac2}.



\begin{thebibliography}{99}

\bibitem{m9711}J.~M.~Maldacena,
``The large $N$ limit of superconformal field theories and supergravity,''
Adv.\ Theor.\ Math.\ Phys.\  {\bf 2}, 231 (1998)
[Int.\ J.\ Theor.\ Phys.\  {\bf 38}, 1113 (1999)]
[arXiv:hep-th/9711200].

\bibitem{m2}
O.~Aharony, S.~S.~Gubser, J.~M.~Maldacena, H.~Ooguri and Y.~Oz,
``Large $N$ field theories, string theory and gravity,''
Phys.\ Rept.\  {\bf 323}, 183 (2000)
[arXiv:hep-th/9905111].

\bibitem{s1}
  G.~Policastro, D.~T.~Son and A.~O.~Starinets,
   ``The shear viscosity of strongly coupled N = 4 supersymmetric Yang-Mills
  plasma,'' Phys.\ Rev.\ Lett.\  {\bf 87}, 081601 (2001)
  [arXiv:hep-th/0104066].


\bibitem{b1}
  A.~Buchel, J.~T.~Liu and A.~O.~Starinets,
   ``Coupling constant dependence of the shear viscosity in N = 4
  supersymmetric Yang-Mills theory,''
  Nucl.\ Phys.\ B {\bf 707}, 56 (2005)
  [arXiv:hep-th/0406264].

\bibitem{b2}
  P.~Benincasa and A.~Buchel,
   ``Transport properties of N = 4 supersymmetric Yang-Mills theory at finite
  coupling,''
  JHEP {\bf 0601}, 103 (2006)
  [arXiv:hep-th/0510041].

\bibitem{s2}
  G.~Policastro, D.~T.~Son and A.~O.~Starinets,
  ``From AdS/CFT correspondence to hydrodynamics. II: Sound waves,''
  JHEP {\bf 0212}, 054 (2002)
  [arXiv:hep-th/0210220].

\bibitem{h1}
  C.~P.~Herzog,
  ``The hydrodynamics of M-theory,''
  JHEP {\bf 0212}, 026 (2002)
  [arXiv:hep-th/0210126].

\bibitem{h2}
  C.~P.~Herzog,
  ``The sound of M-theory,''
  Phys.\ Rev.\ D {\bf 68}, 024013 (2003)
  [arXiv:hep-th/0302086].

\bibitem{b3}
  A.~Buchel,
  ``N = 2* hydrodynamics,''
  Nucl.\ Phys.\ B {\bf 708}, 451 (2005)
  [arXiv:hep-th/0406200].


\bibitem{aby}
  O.~Aharony, A.~Buchel and A.~Yarom,
  ``Holographic renormalization of cascading gauge theories,''
  Phys.\ Rev.\ D {\bf 72}, 066003 (2005)
  [arXiv:hep-th/0506002].

\bibitem{s3}
  A.~Parnachev and A.~Starinets,
  ``The silence of the little strings,''
  JHEP {\bf 0510}, 027 (2005)
  [arXiv:hep-th/0506144].



\bibitem{b4}
  P.~Benincasa, A.~Buchel and A.~O.~Starinets,
  ``Sound waves in strongly coupled non-conformal gauge theory plasma,''
  Nucl.\ Phys.\ B {\bf 733}, 160 (2006)
  [arXiv:hep-th/0507026].


\bibitem{b5}
  A.~Buchel,
  ``A holographic perspective on Gubser-Mitra conjecture,''
  Nucl.\ Phys.\ B {\bf 731}, 109 (2005)
  [arXiv:hep-th/0507275].


\bibitem{b6}
  A.~Buchel,
  ``Transport properties of cascading gauge theories,''
  Phys.\ Rev.\ D {\bf 72}, 106002 (2005)
  [arXiv:hep-th/0509083].

\bibitem{b7}
  P.~Benincasa and A.~Buchel,
  ``Hydrodynamics of Sakai-Sugimoto model in the quenched approximation,''
  Phys.\ Lett.\ B {\bf 640}, 108 (2006)
  [arXiv:hep-th/0605076].


\bibitem{u1}
  A.~Buchel and J.~T.~Liu,
  ``Universality of the shear viscosity in supergravity,''
  Phys.\ Rev.\ Lett.\  {\bf 93}, 090602 (2004)
  [arXiv:hep-th/0311175].


\bibitem{u2}
  P.~Kovtun, D.~T.~Son and A.~O.~Starinets,
   ``Viscosity in strongly interacting quantum field theories from black hole
  physics,''
  Phys.\ Rev.\ Lett.\  {\bf 94}, 111601 (2005)
  [arXiv:hep-th/0405231].


\bibitem{u3}
  A.~Buchel,
   ``On universality of stress-energy tensor correlation functions in
  supergravity,''
  Phys.\ Lett.\ B {\bf 609}, 392 (2005)
  [arXiv:hep-th/0408095].


\bibitem{c1}
  M.~Cvetic {\it et al.},
  ``Embedding AdS black holes in ten and eleven dimensions,''
  Nucl.\ Phys.\ B {\bf 558}, 96 (1999)
  [arXiv:hep-th/9903214].


\bibitem{c2}
  A.~Buchel,
   ``Higher derivative corrections to near-extremal black holes in type IIB
  supergravity,''
  Nucl.\ Phys.\ B {\bf 750}, 45 (2006)
  [arXiv:hep-th/0604167].


\bibitem{stu}
  K.~Behrndt, M.~Cvetic and W.~A.~Sabra,
  ``Non-extreme black holes of five dimensional N = 2 AdS supergravity,''
  Nucl.\ Phys.\ B {\bf 553}, 317 (1999)
  [arXiv:hep-th/9810227].

\bibitem{a1}
  D.~T.~Son and A.~O.~Starinets,
  ``Hydrodynamics of R-charged black holes,''
  JHEP {\bf 0603}, 052 (2006)
  [arXiv:hep-th/0601157].

\bibitem{a2}
  J.~Mas,
  ``Shear viscosity from R-charged AdS black holes,''
  JHEP {\bf 0603}, 016 (2006)
  [arXiv:hep-th/0601144].

\bibitem{a3}
  K.~Maeda, M.~Natsuume and T.~Okamura,
  ``Viscosity of gauge theory plasma with a chemical potential from AdS/CFT,''
  Phys.\ Rev.\ D {\bf 73}, 066013 (2006)
  [arXiv:hep-th/0602010].

\bibitem{a4}
  O.~Saremi,
   ``The viscosity bound conjecture and hydrodynamics of M2-brane theory at
  finite chemical potential,''
  arXiv:hep-th/0601159.


\bibitem{bl}
  A.~Buchel and J.~T.~Liu,
 ``Gauged supergravity from type IIB string theory on Y(p,q) manifolds,''
  arXiv:hep-th/0608002.

\bibitem{y1}
  J.~P.~Gauntlett, D.~Martelli, J.~Sparks and D.~Waldram,
  ``Supersymmetric AdS(5) solutions of M-theory,''
  Class.\ Quant.\ Grav.\  {\bf 21}, 4335 (2004)
  [arXiv:hep-th/0402153].

\bibitem{y2}
  J.~P.~Gauntlett, D.~Martelli, J.~Sparks and D.~Waldram,
  ``Sasaki-Einstein metrics on S(2) x S(3),''
  Adv.\ Theor.\ Math.\ Phys.\  {\bf 8}, 711 (2004)
  [arXiv:hep-th/0403002].

\bibitem{y3}
  D.~Martelli and J.~Sparks,
   ``Toric geometry, Sasaki-Einstein manifolds and a new infinite class of
  AdS/CFT duals,''
  Commun.\ Math.\ Phys.\  {\bf 262}, 51 (2006)
  [arXiv:hep-th/0411238].

\bibitem{y4}
  S.~Benvenuti, S.~Franco, A.~Hanany, D.~Martelli and J.~Sparks,
   ``An infinite family of superconformal quiver gauge theories with
  Sasaki-Einstein duals,''
  JHEP {\bf 0506}, 064 (2005)
  [arXiv:hep-th/0411264].


\bibitem{ss}
  D.~T.~Son and A.~O.~Starinets,
   ``Minkowski-space correlators in AdS/CFT correspondence: Recipe and
  applications,''
  JHEP {\bf 0209}, 042 (2002)
  [arXiv:hep-th/0205051].


\bibitem{sc1}
  S.~R.~Das, G.~W.~Gibbons and S.~D.~Mathur,
 ``Universality of low energy absorption cross sections for black holes,''
  Phys.\ Rev.\ Lett.\  {\bf 78}, 417 (1997)
  [arXiv:hep-th/9609052].


\bibitem{sc2}
  R.~Emparan,
  ``Absorption of scalars by extended objects,''
  Nucl.\ Phys.\ B {\bf 516}, 297 (1998)
  [arXiv:hep-th/9706204].

\bibitem{ne2}
G.~Policastro, D.~T.~Son and A.~O.~Starinets,
``From AdS/CFT correspondence to hydrodynamics,''
JHEP {\bf 0209}, 043 (2002) [arXiv:hep-th/0205052].



\end{thebibliography}
\end{document}